\documentclass[conference]{IEEEtran}
\IEEEoverridecommandlockouts
\ifCLASSINFOpdf
\else
\fi
\usepackage{amsmath}
\usepackage{url} 
\usepackage{graphicx}
\usepackage[acronym,nonumberlist]{glossaries}
\setglossarystyle{list}
\usepackage{verbatim}
\usepackage{orcidlink} 

\newacronym{fi}{FI}{Fault Injection}
\newacronym{iot}{IoT}{Internet of Things}
\newacronym{isa}{ISA}{Instruction Set Architecture}
\newacronym{fia}{FIA}{Fault Injection Attack}
\newacronym{soc}{SoC}{System on Chip}
\newacronym{avf}{AVF}{Architectural Vulnerability Factor}
\newacronym{cip}{CIP}{Candidate Injection Point}
\newacronym{fissc}{FISSC}{Fault Injection and Simulation Secure Collection}

\begin{document}
%
\title{InjectV: Modeling Fault Injection Attacks in RISC-V Simulation Environment
\thanks{
        This work was supported by project COLTRANE-V funded by the Ministero dell'Università e della Ricerca within the PRIN 2022 program (D.D.104 - 02/02/2022).
    }
}
%
%
%

\author{
\IEEEauthorblockN{
Niccolò Lentini\textsuperscript{1}\orcidlink{0009-0008-6639-7414},
Giorgio Fardo\textsuperscript{1,2},
Stefano Di Carlo\textsuperscript{1}\orcidlink{0000-0002-7512-5356},
Alessandro Savino\textsuperscript{1}\orcidlink{0000-0003-0529-7950}
}

\vspace{0.3em}

\IEEEauthorblockA{\textsuperscript{1}Control and Computer Engineering Department, Politecnico di Torino, Turin, Italy}

\IEEEauthorblockA{\textsuperscript{2}Univ. Grenoble Alpes, CEA, List, F-38000 Grenoble, France}

\IEEEauthorblockA{
niccolo.lentini@polito.it, giorgio.fardo@cea.fr, stefano.dicarlo@polito.it, alessandro.savino@polito.it
}
}

%
%

\markboth{}%
{}
%



\maketitle

\begin{abstract}
\glspl{fia} are a significant threat to hardware security, capable of compromising systems by inducing malicious faults in computation or storage. Evaluating resilience against such attacks is challenging due to the high cost, complexity, and limited availability of physical fault experiments, particularly during pre-silicon development. Architectural-level simulation offers a developer-oriented, white-box perspective for systematic vulnerability assessment.
This paper introduces InjectV, a fault injection attack framework for RISC-V platforms built on the gem5 simulator. InjectV enables precise, guided fault injection at security-critical execution points, such as control-flow decisions, counters, and comparisons, allowing systematic exploration of attack vectors. It currently supports transient fault attacks in registers and memory, broadening its ability to simulate diverse attack scenarios.
Experimental results on security benchmarks from the FISSC suite, including hardened variants of the VerifyPIN application, demonstrate InjectV's ability to effectively identify fault-injection points, achieving a 95.8\% time-saving advantage over traditional fault injection approaches.
\end{abstract}


%
\IEEEpeerreviewmaketitle

\glsresetall

\section{Introduction} \label{intro}

Nowadays, \glspl{fia} are a well-established class of hardware security threats where carefully timed and localized faults can be used to corrupt control flow, bypass authentication checks, or extract sensitive information from otherwise protected systems \cite{Gangolli2022,Boulifa2025,Portolan:2019,Ravi2024}. 
Unlike random faults traditionally studied in reliability analysis, \glspl{fia} are intentional and targeted. Attackers exploit specific execution points, where a single transient fault can invalidate a protection mechanism or alter program semantics. This shifts the focus of \gls{fia} resilience analysis, making the challenge of restricting and guiding the analysis to those trigger points rather than exhaustively exploring arbitrary fault locations.

Evaluating resilience against \glspl{fia} on real hardware is expensive, difficult to reproduce, and often infeasible during early design phases. Moreover, exhaustive exploration of the fault space on physical devices is intractable due to the combinatorial explosion of fault parameters, including timing, location, and target state. As a result, developers lack practical tools to systematically assess fault vectors and software countermeasures before deployment \cite{murdock2020plundervolt}.

Simulation-based fault injection offers a cost-effective, repeatable alternative to physical testing, enabling scalable security evaluation during early design. However, existing literature lacks tools specifically optimized for large-scale attack-oriented \gls{fia} analysis.

This paper presents \textbf{InjectV}, a gem5-based \cite{gem5} RISC-V fault-injection framework that combines preprocessing-driven identification of security-relevant execution points with guided register- and memory-fault campaigns. RISC-V is targeted due to its increasing adoption in embedded and security-critical systems \cite{Werneretal2019,Gerlachetal2023,Andersetal2023,Farnaghinejad:2025aa}, while gem5 provides a modular full-system simulation infrastructure that can be extended to other instruction set architectures. Experimental results on FISSC VerifyPIN benchmarks show that InjectV enables scalable attack-oriented exploration and achieves a 24$\times$ efficiency gain over random injection.


\section{Background}\label{background}

A \gls{fia} is a physical attack in which an adversary deliberately induces faults to perturb hardware components such as registers, memory, or control logic during program execution, to force the system to deviate from its intended operation. \glspl{fia} violate security through transient corruptions, as early research demonstrates that even minimal, targeted faults can fully compromise secret material \cite{boneh1997,kocher1999,Joye2012,Teixeira2019}.

Since most faults are masked or cause benign crashes \cite{MoradiFailureIdentify}, effective evaluation requires identifying specific software–architecture contexts where perturbations become exploitable behaviors. To achieve this, analysis must be system-aware and emulate the full execution stack. We therefore utilize the gem5 full-system simulator \cite{gem5} to provide the precise, reproducible control necessary to study complex attack vectors across the hardware–software interface.

%


\section{Related Works}
\label{relwork}

Existing fault-injection tools provide fine-grained control over fault models, but offer limited support for relating faults to security-relevant program semantics \cite{Alonso:2024,Magliano:2025aa}. They therefore emphasize broad fault-space coverage rather than attacker-guided injections at critical branches, comparisons, or counters.

Physical fault-injection studies provide realistic evidence of attack feasibility. Dutertre et al. \cite{dutertre2021emskip} showed that electromagnetic injection can induce multiple consecutive instruction skips and evaluated the resulting model on a PIN-code verification algorithm. Their work also led to the \gls{fissc} benchmark suite \cite{FISCC}, which provides hardened program variants with explicit security objectives. Complementary software-level approaches, such as the formal models proposed by Moro et al. \cite{moro2014formal}, reason about instruction skips and control-flow perturbations to derive and assess countermeasures.

Simulation and emulation frameworks improve scalability and reproducibility. Within gem5, tools such as GemFI \cite{gemfi}, gemV-tool \cite{gemvtool}, and gem5-MARVEL \cite{marvel} support large-scale fault-injection campaigns and detailed architectural or microarchitectural resilience studies. However, these frameworks primarily target reliability-oriented analysis, focusing on coverage, fault vulnerability, and dependability metrics rather than adversarial strategies and workload-level security outcomes. QEMU-based approaches \cite{Bekele:2023} and binary-level tools such as FaultFinder \cite{Murdock2024} offer faster exploration, but trade off detailed timing fidelity, microarchitectural modeling, or full-system visibility.
InjectV addresses this gap by combining gem5 full-system simulation with preprocessing-driven identification of potential attack trigger points. Unlike approaches centered on uniform fault-space exploration, InjectV guides register and memory fault campaigns toward security-relevant execution points while preserving reproducible scheduling and end-to-end analysis across the hardware-software stack.

\section{Methodology} \label{method}

InjectV is structured around a systematic preprocessing stage that isolates potentially exploitable execution windows and guides the following fault-injection campaign. The framework operates on the architectural state within a full-system RISC-V gem5 simulation. 
Therefore, the methodology is not tied to a specific core implementation, although the available microarchitectural detail depends on the selected gem5 CPU model. 

\begin{figure}[htb]
    \centering
    \includegraphics[width=0.98\linewidth]{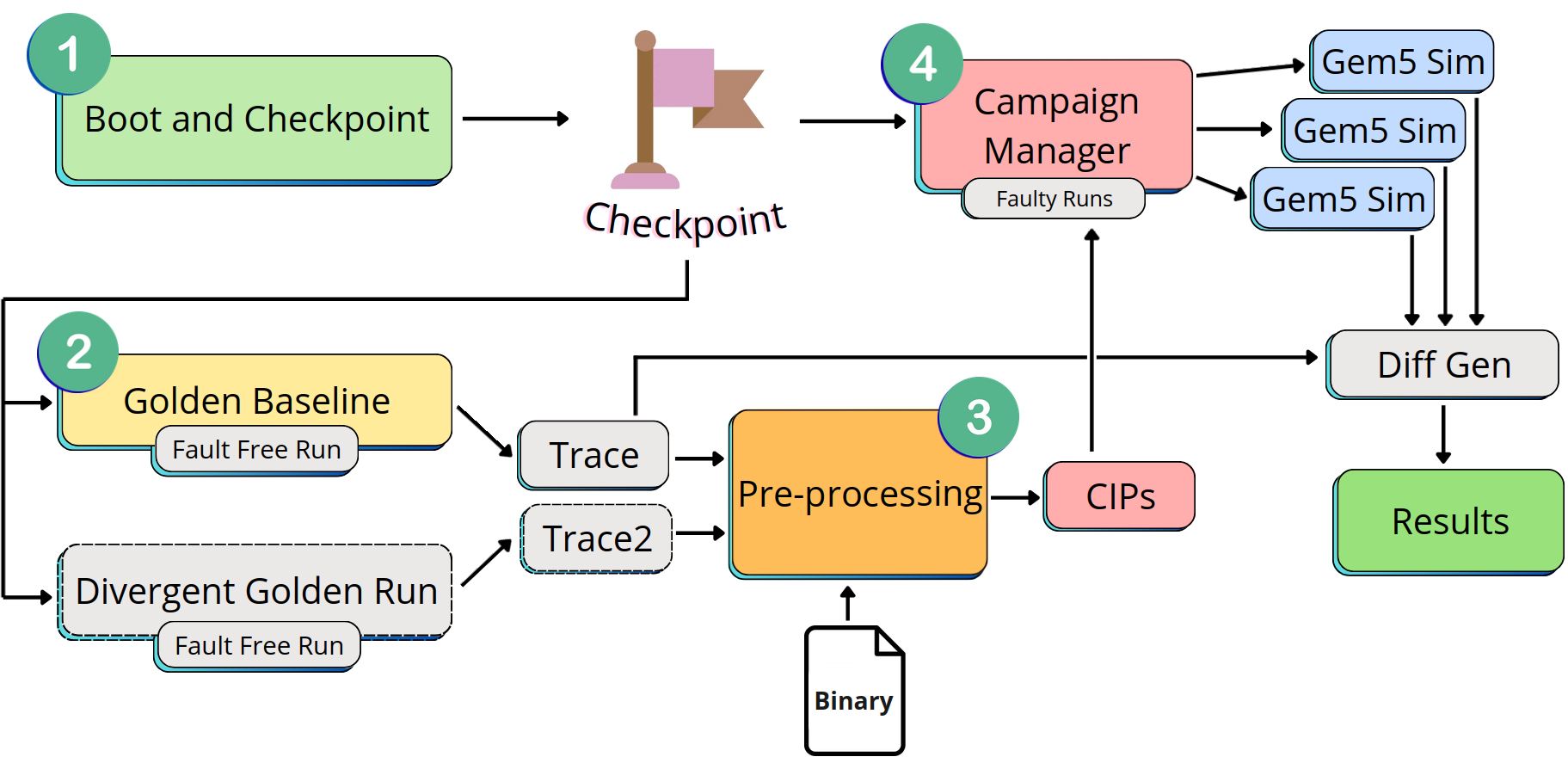}
    \caption{Execution workflow of the proposed fault injection attack framework.}
    \label{fig:workflow_revised}
\end{figure}

Figure \ref{fig:workflow_revised} illustrates the execution flow of the fault injection attack campaign. The overall workflow consists of four main steps:

1) \textbf{Preparation phase:} the system is booted once, and a checkpoint is created after the operating system completes initialization. Subsequent fault-injection experiments start directly from this checkpoint, avoiding repeated boot, reducing trace size, and simplifying simulation and result collection.

2) \textbf{Golden and Divergent baseline establishment:} a fault-free execution trace, called the \textit{golden baseline}, is generated to represent the normal application execution.\footnote{A trace denotes the ordered sequence of executed instructions collected during gem5 simulation, including simulation tick, program counter, disassembled instruction, accessed architectural registers, and, when available, memory access information.} Optionally, one or more \textit{divergent baselines} are collected to represent legitimate executions following alternative control-flow paths, e.g., verification success against failure. These traces are then provided to the preprocessing phase.

3) \textbf{Preprocessing phase:} this phase represents the key novelty of the approach. It correlates the execution traces with the target binary to identify security-relevant execution points and to generate a reduced set of candidate \glspl{cip}.

4) \textbf{Fault injection phase:} this phase is managed by a campaign manager, which parallelizes the gem5 simulations, parses and activates the faults specified in the fault files generated in the previous step, classifies the outcome of each fault-injection run, generates the differences with respect to the golden trace, and produces the final report.

\textbf{Threat and Attack Model:} The framework assumes a realistic physical attacker with local access to the device, aiming to bypass security mechanisms using techniques such as laser or voltage/clock glitching.

InjectV abstracts these effects as deterministic, transient bit-level corruptions injected at a specific simulation tick, targeting either CPU architectural registers or physical memory words. This models the software-visible behavior of modern physical fault attacks.

\textbf{Preprocessing and Candidate Injection Points (CIP) Generation:} As highlighted in Section \ref{background}, untargeted injection yields predominantly masked or non-actionable faults. To explicitly reduce the exploration space, the preprocessing stage produces a list of \glspl{cip}.

The framework cross-analyzes the execution trace (and a divergent trace, when provided) with the target ELF binary to systematically identify security-relevant execution points based on the following architectural characteristics:

\textbf{Control-Flow and Branch Behavior:} The framework parses the assembly code extracted from the target binary file to identify conditional branch instructions. Combining this information with the execution trace provides the simulation ticks at which instructions are executed, allowing the framework to determine whether a branch was taken and to locate critical branch evaluations together with their temporal injection window.

\textbf{Data Dependencies and Write-Before-Use:} InjectV analyzes register data liveness by identifying write-before-use patterns directly from the execution trace and binary image. For each register write, the tool determines how many subsequent instructions are executed before the value is first consumed. If this instruction distance falls within a user-defined window, the location is considered a valid candidate injection point.

\textbf{Differential Execution Divergence:} When a divergent execution trace is provided, the framework compares it with the golden trace to identify the first diverging execution point and correlates it with the corresponding instruction and simulation tick, e.g., valid versus invalid PIN code verification.

\textbf{Instruction-Level Susceptibility and Logic Flips:} The framework analyzes the opcodes of instructions extracted from the binary and observed in the execution trace to identify operations whose semantics can be altered through bit-level perturbations. Candidate instructions are evaluated according to the fault model parameters, including the maximum number of bits that may flip and whether flipped bits must be contiguous. Instructions requiring fewer flipped bits and contiguous bit patterns are ranked as better candidates, enabling the identification of instruction logic flips, e.g., \texttt{beq} $\leftrightarrow$ \texttt{bne}, or transformations that convert security-relevant instructions into NOPs. This process outputs a JSON manifest of \glspl{cip}, each uniquely identified by a temporal coordinate (i.e., the simulation tick), a spatial target (either a physical memory address or a CPU register), and a viable fault model.

\begin{figure}[!t]
    \centering
    \includegraphics[width=0.84\linewidth]{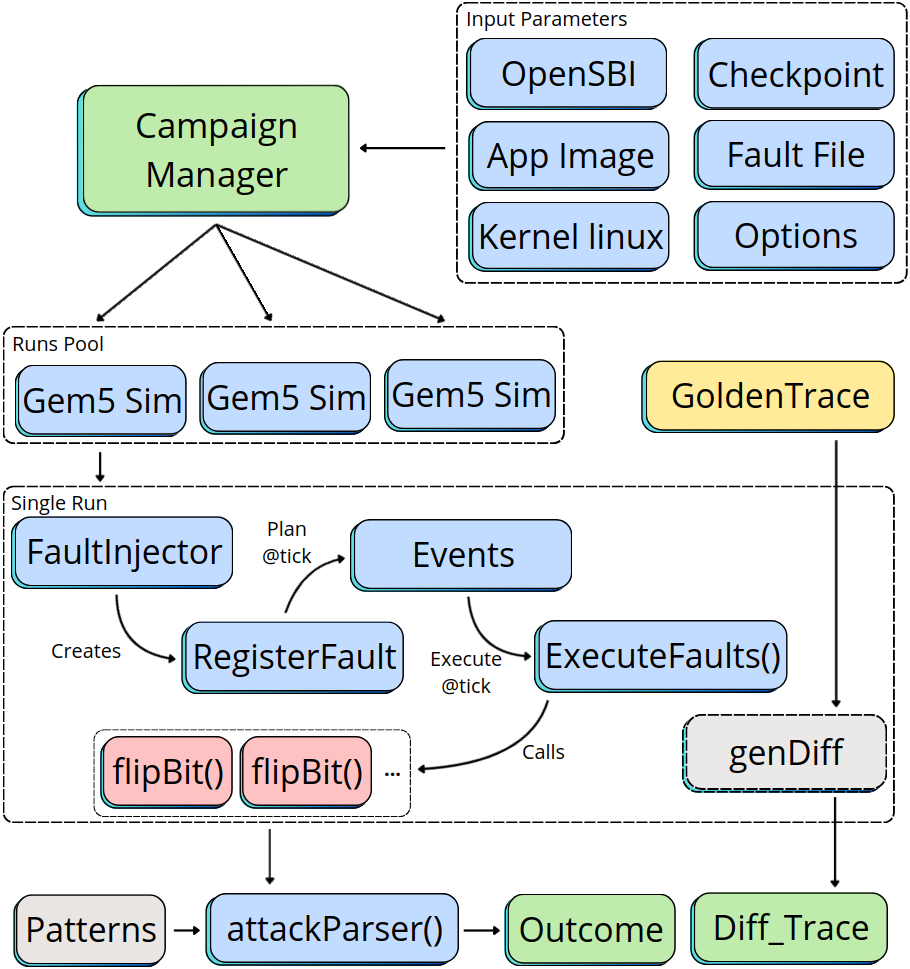}
    \caption{Campaign Manager workflow.}
    \label{fig:campaignManager}
    \vspace{-4mm}
\end{figure}

\textbf{Guided Injection Strategy:} The generated \gls{cip} manifest is used by the framework to craft targeted fault injections automatically. The campaign manager then uses the resulting fault list to schedule each injection in a dedicated gem5 simulation as part of a precise campaign.

\textbf{Scalable Campaign Orchestration and Outcome Classification:} InjectV decouples the fault-injection mechanics from an automated \textit{Campaign Management} layer. Rather than repeatedly restarting the evaluation environment, InjectV groups experiments into discrete, independent campaigns that can be run in parallel. The Campaign Manager launches multiple isolated gem5 instances that resume from a pre-computed stable checkpoint, delivering \textbf{zero OS boot overhead} and improving throughput.

Eventually, InjectV includes an attack-oriented classification stage that separates system-level execution issues from application-level outcomes resulting from the executed workload. \textbf{Host-Level Supervision (TIMEOUT):} runs exceeding a timeout derived from the golden, fault-free, execution time are terminated by the host and classified as \textbf{TIMEOUT}.  \textbf{Target-Level Outcome Classification (CRASH, NO\_EFFECT, UNKNOWN):} runs that finish within the timeout are classified by parsing guest-visible telemetry like terminal output and explicit failure markers. 

Clear OS/application failures, such as kernel panic, are labeled \textbf{CRASH}. Runs whose outputs match the golden baseline are labeled \textbf{NO\_EFFECT}. If the observed output does not match any expected pattern, the run is labeled \textbf{UNKNOWN} and flagged for follow-up analysis. 

Finally, an execution is labeled \textbf{SUCCESS} \emph{only} when the injected perturbation satisfies a user-defined, application-dependent security predicate.

\section{Results} \label{results}

In this section, we describe the experimental campaign used to evaluate InjectV.

\texttt{VerifyPIN} is the benchmark workload used to evaluate InjectV on a realistic, security-relevant control flow. It implements a PIN verification routine that compares a user-provided PIN against a stored reference, while enforcing authentication checks and limiting attempts. \texttt{VerifyPIN} is part of the \gls{fissc} \cite{FISCC}, a suite of C programs designed to study fault-injection attacks on applications hardened by means of countermeasures, such as Hardened booleans, Backup copy, Double test, Step counter, and programming features, such as Inlined calls and Fixed time loop. \gls{fissc} provides several \texttt{VerifyPIN} variants with progressively increasing levels of software hardening.

This setup enables a controlled comparison between an unprotected implementation and progressively hardened verification flows under identical simulation conditions. For each \texttt{VerifyPIN} version, the InjectV preprocessing stage produced, on average, approximately 5,000 candidate fault injections. These were executed in a large-scale simulation campaign and classified into workload-driven outcome categories.

To properly classify \textbf{SUCCESS}, the user-defined predicate parses the final trace status line, which reports three \texttt{VerifyPIN} variables: countermeasure activation, authentication success, and the remaining PIN attempts. A run is classified as \textbf{SUCCESS} when the final state indicates authentication bypass, countermeasure-triggered authentication, or a reset of the protection counter enabling further attempts. Runs that deviate from the golden execution but do not satisfy this predicate are labeled \textbf{PARTIAL\_SUCCESS}. This class captures application-visible deviations that are not directly exploitable according to the selected security objective. This predicate-based definition distinguishes generic divergence from exploitability and allows campaign outcomes to be interpreted in terms of security impact.

To assess the guided preprocessing stage, we compared it with random exploration campaigns that uniformly sample the fault space in time and space. To ensure a fair comparison, random campaigns used the same number of injections and the same register/memory fault mode distribution as the guided ones. Overall, 46,067 $\times$ 2 simulations were executed on 37 cores, with 33.6GB peak RAM usage, 30 seconds average runtime per simulation, and around 20 hours for diff generation.

\begin{figure}[tb]
    \centering
    \includegraphics[width=0.98\linewidth]{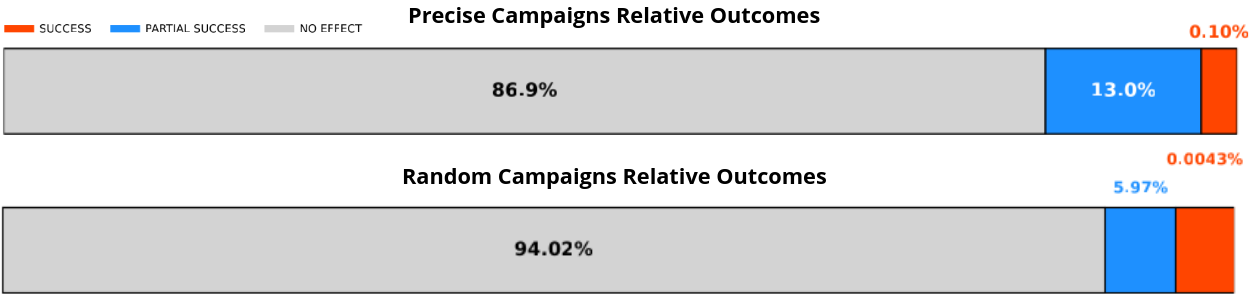}
    \caption{Relative Outcomes Distribution.}
    \label{fig:relatives}
\end{figure}

\begin{figure}[tb]
    \centering
    \includegraphics[width=0.98\linewidth]{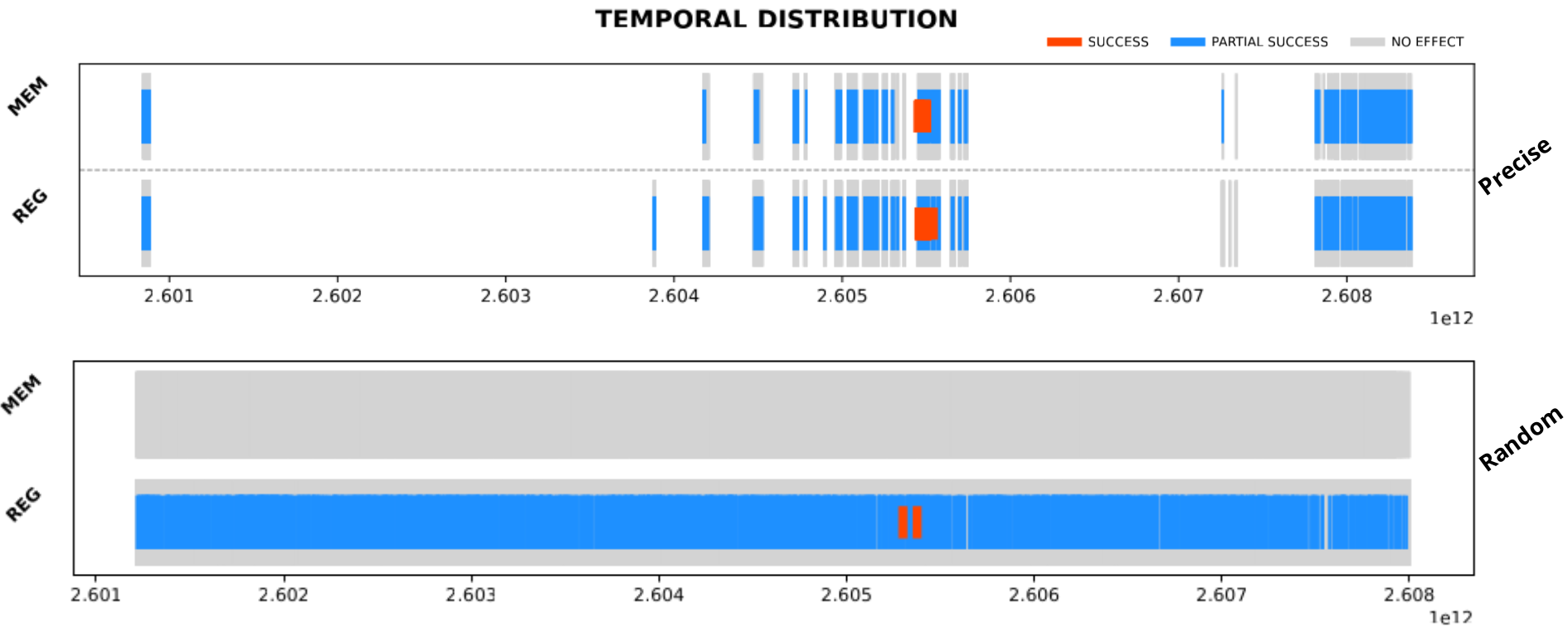}
    \vspace{-2mm}
    \caption{Temporal distribution of injections for guided and random campaigns.}
    \label{fig:temporal}
    \vspace{-2mm}
\end{figure}

\textbf{Guided vs Random Fault Exploration:} The results of the guided and random campaigns were analyzed across the temporal distribution of injections, the spatial distribution across memory addresses, and the spatial distribution across architectural registers. For each injection, outcomes were categorized consistently with the scheme defined in Section \ref{method}.
The campaigns confirm the effectiveness of the guided approach. As reported in Fig. \ref{fig:relatives}, InjectV discovered 48 successful security violations, 27 from memory corruption and 21 from register perturbation, whereas random injection discovered only 2.
The temporal analysis in Fig. \ref{fig:temporal} highlights a clear difference between guided and random exploration. Guided injections cluster around execution regions where the PIN value is processed, and the verification logic executes, while random exploration distributes injections uniformly across the execution timeline, producing mostly \textit{NO\_EFFECT} outcomes.

\begin{figure}[tb]
    \centering
    \includegraphics[width=0.98\linewidth]{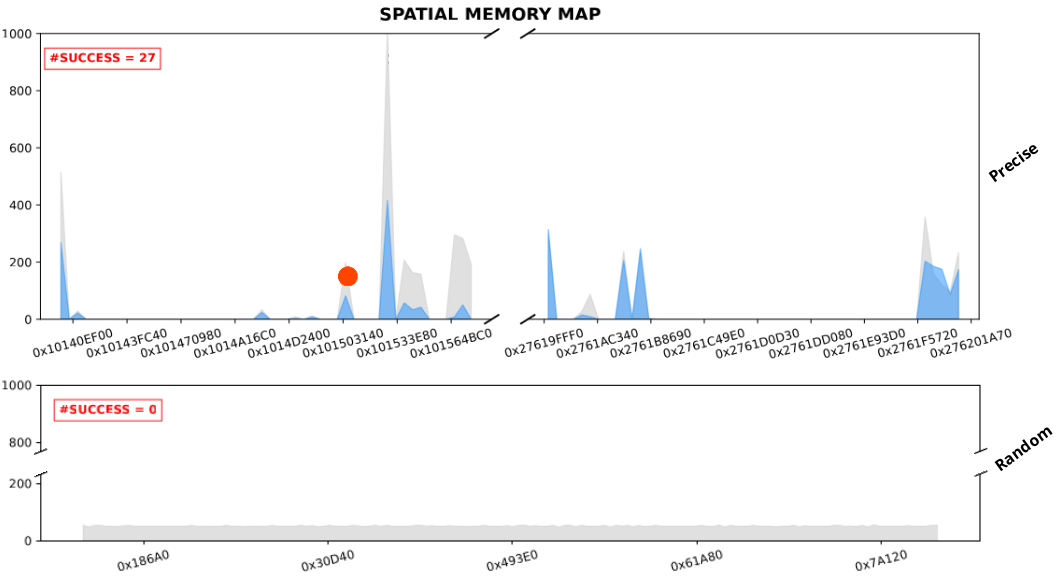}
    \caption{Spatial Distribution of memory injections.}
    \label{fig:spatmem}
\end{figure}

\begin{figure}[tb]
    \centering
    \includegraphics[width=0.98\linewidth]{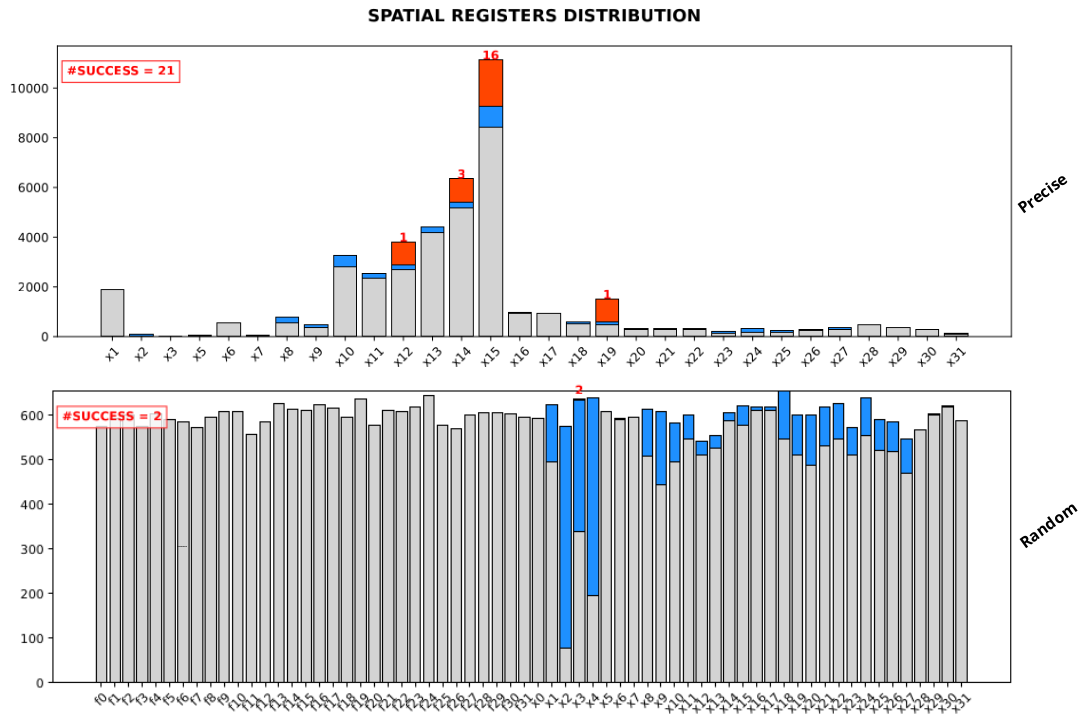}
    \caption{Spatial Distribution of register injections.}
    \label{fig:spatreg}
\end{figure}

Figures \ref{fig:spatmem} and \ref{fig:spatreg} show that guided injections concentrate on a restricted set of sensitive memory regions and registers. In particular, all successful memory corruptions fall along the PIN verification path, while register \texttt{x15} accounts for 16 of the 21 successful register-based violations. Conversely, random exploration uniformly targets the memory space and register file, producing mainly \textit{NO\_EFFECT} outcomes.

Across the different \texttt{VerifyPIN} versions, the number of successful attacks remains similar. This suggests that the guided preprocessing stage often identifies control-flow decisions that directly govern authentication, allowing faults to bypass higher-level countermeasures. This motivates future work on attack strategies targeting specific software protections.

Eventually, the success rate comparison between random and guided campaigns reveals a significant disparity in efficiency. To achieve the same success rate as the guided campaign, a random approach would require an estimated 1,104,000 injections, representing a 24$\times$ increase in computational effort. Consequently, the preprocessing methodology presented in this paper achieves a 95.8\% time-saving advantage over a brute-force random injection strategy.

\section{Conclusion} \label{concl}
This paper introduced InjectV, an attack-centric fault-injection framework for full-system RISC-V simulation built on gem5. By using preprocessing to identify security-relevant execution points, InjectV reduces the search space for exploitable vulnerabilities and achieves a 24$\times$ improvement in analysis efficiency over random injection. Future work will extend the framework to multi-fault scenarios and broader attack models.

To support open science in research, InjectV is publicly available at the official GitHub repository: \url{https://github.com/smilies-polito/injectv}.


%

\ifCLASSOPTIONcaptionsoff
  \newpage
\fi



%
\newpage
\bibliographystyle{IEEEtran}
\bibliography{references}

%






\end{document}